\documentstyle[12pt,aasms4]{article}

\def	\Angstrom {\,{\rm\AA}}
\def	\AU	{\,{\rm AU}}
\def	\beq	{\begin{equation}}
\def	\cm	{\,{\rm cm}}

\def	\eeq	{\end{equation}}

\def	\g	{\,{\rm g}}

\def	\kpc	{\,{\rm kpc}}

\def	\Msol	{M_{\odot}}
\def	\micron	{\,\mu{\rm m}}

\begin{document}

\title{Constraints on Cold ${\rm H}_2$ Clouds from Gravitational 
Microlensing Searches.}
\author{R.R.Rafikov\altaffilmark{1} \and B.T.Draine\altaffilmark{1}}
\altaffiltext{1}{Princeton University Observatory, Princeton, NJ 08544}

\begin{abstract}
It has been proposed that the Galaxy might contain a population of 
cold clouds in numbers sufficient to account for a substantial 
fraction of the total mass of the Galaxy. These clouds would have  
masses of  $\sim  10^{-3} M_{\odot}$
and sizes $\sim 10$ AU. We consider here the lensing effects of such
clouds  on the light from background stars. A semianalytical formalism 
for calculation of the magnification
 event rate produced by such gaseous lensing 
is developed, taking into account the spatial distribution of the dark
matter in the Galaxy, the velocity distribution of the lensing clouds 
and source stars, and  motion of the observer. Event rates
are calculated for the case of gaseous lensing of stars in the Large 
Magellanic Cloud and results are directly compared with the 
results of the search for gravitational microlensing events 
undertaken by the MACHO collaboration.
The MACHO experiment strongly constrains the properties of the proposed 
molecular clouds, but does not completely rule them out. 
Future monitoring programs will either detect or more strongly constrain 
this proposed population.
\end{abstract}

\keywords{galaxies: halos --- galaxies: ISM --- Galaxy: halo
--- gravitational lensing --- ISM: clouds}

\section{Introduction}
The nature 
of the dark matter dominating  the mass 
of the Galaxy remains elusive. Although  nonbaryonic forms 
of dark matter are one  possible explanation, it is possible
that much or most of the mass may be baryonic.
Recent results from gravitational microlensing experiments, such
as MACHO and EROS (Alcock et al. 2000;
Lasserre et al. 2000), have put  severe constraints on the 
abundances of  compact baryonic objects in the Galaxy, - 
such as brown dwarfs, old white dwarfs, neutron stars, or black hole 
remnants - thus virtually 
removing stellar objects as  candidates for  being the major 
constituent of the dark matter (Alcock et al. 2000; 
Freese, Fields, \& Graff 2000;
Lasserre et al. 2000).   

It has recently been proposed that  the dark matter in  the Galaxy
could consist of small, cold clouds of H$_2$
(Pfenniger, Combes, \& Martinet 1994; De Paolis et al. 1995a, 1995b, 1996;
Gerhard \& Silk 1996). Walker \& Wardle (1998) pointed out that
such clouds could be responsible for the extreme scattering events 
(ESEs; Fiedler et al. 1994), which would occur when the line of sight to
an extragalactic 
point radio source  is crossed by the photoionized cloud envelope. 
In a later paper, Wardle \& Walker (1999) considered the 
 thermal stability of such clouds and showed that these clouds
could be stable against heating by  cosmic rays. 
These clouds could have masses of the order of Jupiter's mass
($10^{-3} M_{\odot}$) and radii $ R_{cl} \approx 10$ AU.
If they exist, these clouds could  also naturally explain the core 
radii of galaxies (Walker 1999) as well as the $\gamma$-ray emission 
from  the Galactic halo (Kalberla, Shchekinov, \& Dettmar 1999; Sciama 2000). 
Cloud-cloud collisions would steadily resupply the Galactic disk with 
gas to sustain steady star formation.

The proposed  clouds are not compact enough to 
produce any gravitational microlensing
(Henriksen \& Widrow 1995)  but 
Draine (1998) demonstrated that
they can produce magnification of background stellar sources. 
Refraction of light passing through the clouds 
 would cause amplification of background
stars  in a way resembling gravitational microlensing,
which provides us with the possibility of using searches 
for  gravitational microlensing events to either detect such gas clouds
or constrain the properties of the cloud population. 
 
In this paper we calculate the lensing 
rate of the stars in the Large Magellanic Cloud (LMC) by clouds of
molecular hydrogen, taking into account the spatial distribution  
of the  clouds  and all the relevant motions
which contribute to this rate: motion of the observer on the Earth,
proper motion of the LMC, the velocity
distribution of stars in the Cloud, and the velocity distribution 
of lensing clouds in the Galactic halo. We suppose that the dark
matter is composed of clouds of only one size and mass and that the clouds 
are transparent.

The paper is organized as follows:
in \S \ref{Phys} we review some aspects of the lensing by gaseous 
clouds. In \S \ref{sec:rate} we derive basic formulae for 
the rate of lensing events produced by the gaseous clouds and
for timescale distribution of this sort of lensing. In 
\S \ref{results} we present our results in the form of 
parametric plots, covering a wide range of cloud models, and  
compare our results with those obtained by collaborations undertaking
searches for  gravitational microlensing events. We 
determine the region of parameter space  which
is not excluded by these experiments and other constraints.
Finally, in \S \ref{disc}
we compare our results with the limits  placed on cloud models by
other authors from different arguments.

\section{Physics of gaseous lensing\label{sec:phys_of_gas_lens}}\label{Phys}

The physics of lensing by  gaseous clouds  has been discussed
by Draine (1998) and we  repeat here only the salient points.
The refractive index $m$ is related to the gas density $\rho$ by 
\begin{equation}
m(\lambda) = 1 + \alpha(\lambda)\rho ~~~;
\label{refr_ind}
\end{equation}
$\alpha(4400\Angstrom)= 1.243 \cm^3\g^{-1}$ and
$\alpha(6700\Angstrom)= 1.214 \cm^3\g^{-1}$
for H$_2$/He gas with 24\% He by mass 
(AIP Handbook 1972). 

For small deflections, 
a light ray with impact parameter $b$ will be deflected through
an angle
\begin{equation}
\phi(b) = -2\alpha b \int\limits_b^\infty {dr \over (r^2-b^2)^{1/2}}
{d\rho\over dr}  ~~~.
\label{eq:phi}
\end{equation}

\begin{figure}
\vspace{18.cm}
\includegraphics{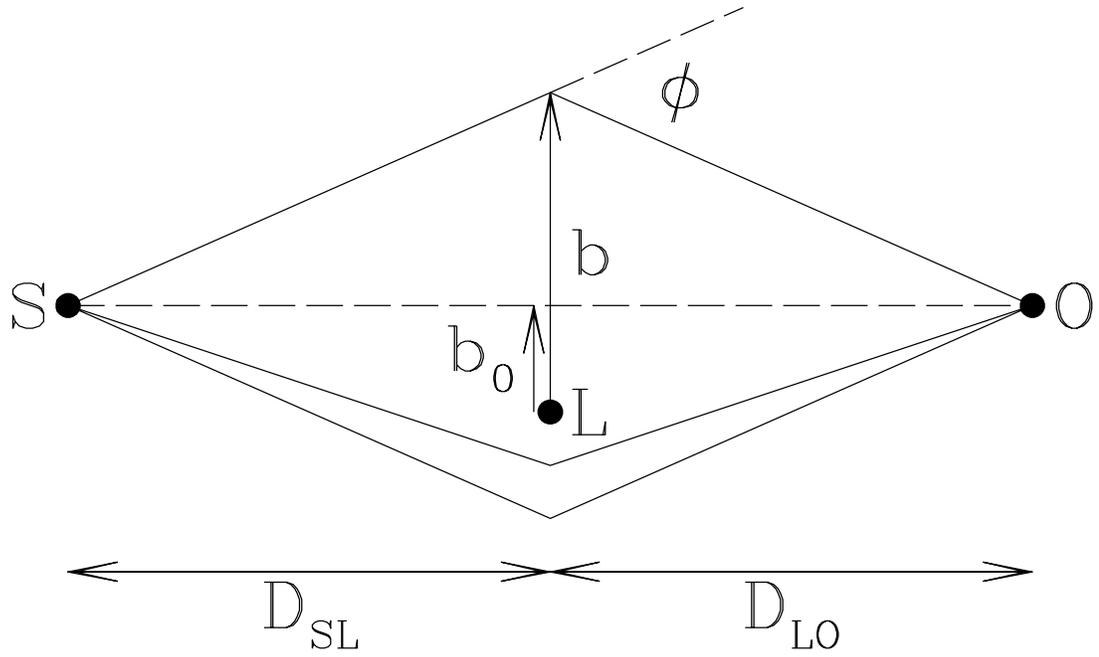}
\caption{
Geometry of lensing. S is the source, the center of the lens is
at L, and O is the observer.
}
\label{picture}
\end{figure}

Let $D_{\rm sl}$  and $D_{\rm ol}$ be 
the distance from source to lens, and
from observer  to lens (see Figure \ref{picture}).
If $b_0$ is the distance of the lens center from the 
straight line from source
to observer, 
then the apparent distance $b$ of the image from the
lens is given by the lensing equation
\begin{equation}
b - b_0 = D\phi(b) ~~~,~~~
D \equiv {D_{\rm sl}D_{\rm ol}\over D_{\rm sl}+D_{\rm ol}} ~~~.
\label{eq:lensing}
\end{equation}

For a point source,
the image magnification is given by
\begin{equation}
M(b) = {|b|\over b_0}{1 \over 1-D\phi^\prime(b)}  ~~~,
\label{eq:mag}
\end{equation}
\begin{equation}
\phi^\prime(b) \equiv {d\phi(b)\over db} =  -2\alpha\int\limits_0^\infty dz 
\left[
{b^2\over r^2}{d^2\rho\over dr^2} + {z^2\over r^3}{d\rho\over dr}
\right] ~~~.
\label{dphidb}
\end{equation}
where $r^2=b^2+z^2$.
For a given $b_0$ 
there will be an odd number $N(b_0)$ of solutions $b_i(b_0)$, $i=1,...,N$.
The total 
amplification $A(b_0)=\sum_{i=1}^N M(b_i)$.
The ``trajectory'' of the lens relative to the source is characterized by
a ``source impact parameter'' $p$ and a displacement $x$ along the
trajectory; for any $x$ we have $b_0=(p^2+x^2)^{1/2}$, and the
``light curve'' is just $A(b_0)$ {\it vs.} $x$.

Following Draine (1998) we define a dimensionless
``strength'' parameter
\begin{equation}
S \equiv {\alpha\langle\rho\rangle D \over R_{cl}} = 
0.355\left({M_{cl}\over10^{-3}\Msol}\right)\left({\AU\over R_{cl}}\right)^4
\left({D\over 10\kpc}\right) ~~~,
\label{eq:Sdef}
\end{equation}
where $M_{cl}, R_{cl}$, and $\langle\rho\rangle\equiv 3 M_{cl}/4\pi R_{cl}^3$ 
are the cloud mass, radius, and mean density.

As the pressure density-relation in such clouds is uncertain, Draine (1998)
considered polytropic equations of state for polytropic index 
$1.5\le n < 5$. (For $n=1.5$ the cloud is isentropic, while for 
$n\to 5 $ the central density becomes infinite).

\begin{figure}
\vspace{18.cm}
\includegraphics{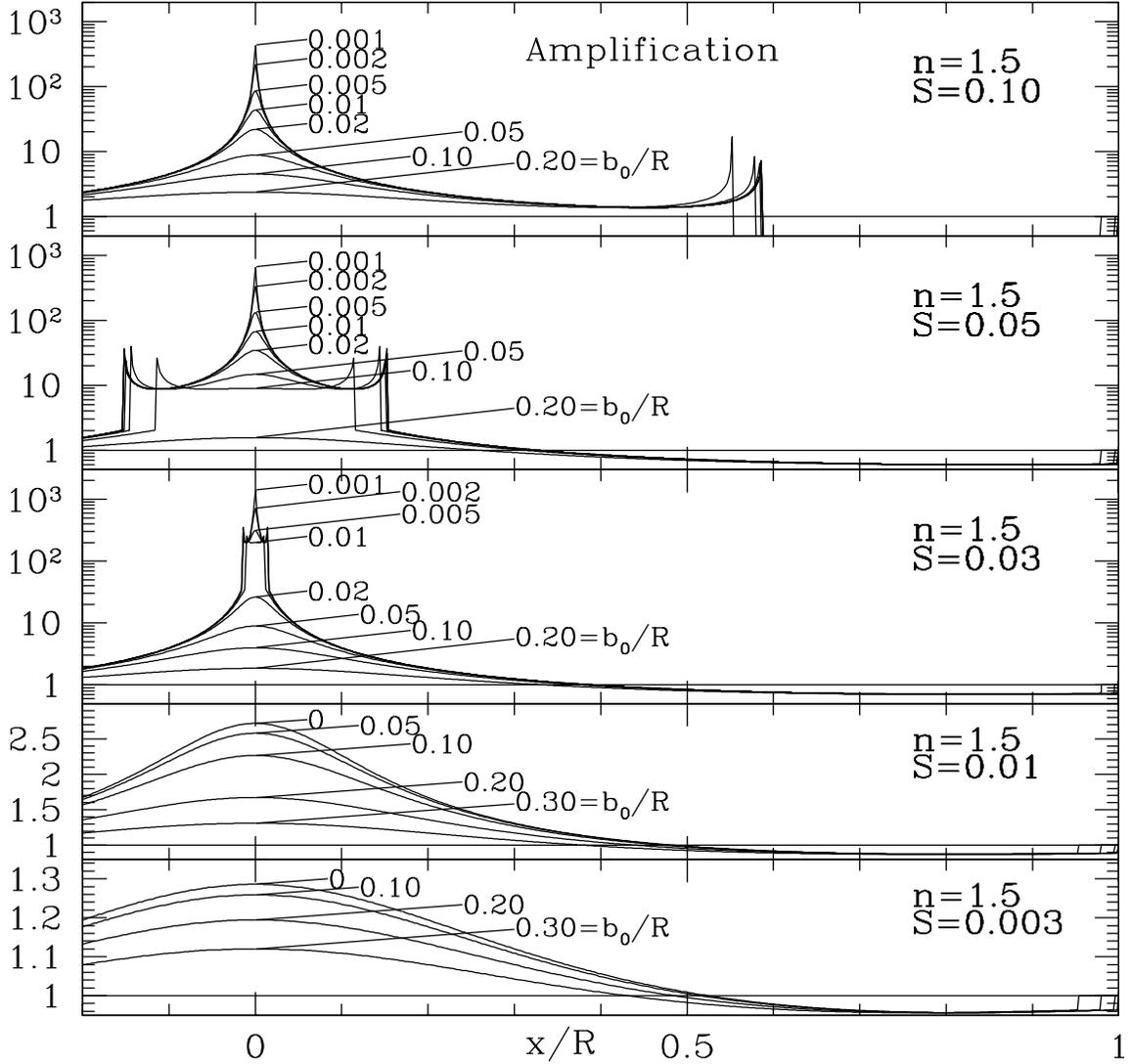}
\caption{
Plots, showing lightcurves for $n=1.5$ with different ratios $S/S_{cr}$:
(from top to bottom) 4, 2, 1.2, 0.4, 0.12. On the three top panels
one can easily see caustics produced by multiple images. For
lightcurves corresponding to other values of $n$ and $S/S_{cr}$
see Draine (1998).
}
\label{lightcurve}
\end{figure}

For each polytropic index $n$ there exists a specific value of 
$S=S_{cr}$ such that for $S<S_{cr}$  equation (\ref{eq:lensing})
has only one solution and magnification of the lens is always finite,
even for $b_0=0$.
 For $S>S_{cr}$ equation (\ref{eq:lensing})
can have three solutions for sufficiently small $b_0$ 
and  caustic lensing can occur. 
In this caustic regime the magnification becomes infinite 
for $b_0=0$ and also for some finite $b_{0c}$ given by the 
condition that at this $b_{0c}$ two of the roots 
of  equation (\ref{eq:lensing})
coincide. At $b_0=b_{0c}$ light curves exhibit conspicuous caustics
with magnification going to infinity. Cases of caustic and 
noncaustic lensing are illustrated in Figure \ref{lightcurve}
for $n=1.5$.

For $S<S_{cr}$, we can obtain  a simple analytical formula describing the
dependence of the central magnification
$M(0)$ upon the strength parameter $S$. If we define dimensionless 
variables $\hat\rho\equiv\rho/\langle\rho\rangle$ and
$\hat r\equiv r/R_{cl}$, it is straightforward to show that
as $b_0\to 0$
\begin{equation}
b - b_0 \to - 2 S b \int\limits_0^1 \frac{d\hat r}{\hat r}. 
{d\hat\rho\over d\hat r}
\label{bofb0}
\end{equation}
The integral in this expression depends only on 
the  cloud density profile.

Substituting this result into (\ref{eq:mag}) we get
\begin{equation}
M(0)=\frac{1}{\left(1-S/S_{cr}\right)^2},
\label{M0}
\end{equation}
where 
\begin{equation}
S_{cr}\equiv - \left(2 \int\limits_0^1 \frac{d\hat r}{\hat r} 
{d\hat\rho\over d\hat r}\right)^{-1}.
\label{s-cr}
\end{equation}
 
Thus $M(0)\to \infty$ as $S\to S_{cr}$, but it is finite for
smaller $S$. At $S=S_{cr}$ the 
number of solutions of equation (\ref{eq:lensing})
 changes from $1$ to $3$ (for $b_0=0$) and for  $S\ge S_{cr}$ we 
enter the caustic regime. 
For $n=1.5$ $S_{cr}=0.026$ and for
$n=4.5$ $S_{cr}=1.72 \times 10^{-6}$.

\section{Rate of gaseous lensing events}\label{sec:rate}

When a cloud crosses the line of sight between the 
distant star and the observer on the Earth, it 
amplifies or attenuates the brightness of the star [in fact each 
gaseous lensing curve exhibits some demagnification before and after 
the peak magnification (Draine 1998), but we will be primarily 
interested in amplification in this paper].
This effect can be observed by searches for 
 gravitational microlensing in our Galaxy
which monitor the brightness of  a large number 
of stars. 

Let  $D_{S}$ be the distance to the
 source stars being lensed by gaseous clouds in the Galaxy.
Consider a lens  located at some distance $D_{OL}=x D_{S}$ 
from the observer;
 the distance from the lens to the source star is  
$D_{LS}=(1-x) D_{S}$, and $D=x(1-x)D_S$.

Both lensing regimes, caustic and noncaustic, are important.
Since the  lensing parameter 
\begin{equation}
S = {\alpha\langle\rho\rangle D_S \over R_{cl}} x (1-x),
\label{eq:Sprpto}
\end{equation}
 maximum  $S$ is attained 
when the lens is placed midway 
between the source and the observer,  and is equal to 
$S_{max}=\alpha\langle\rho\rangle D_S / (4  R_{cl})$. As $x$ changes towards
$x=0$ or $x=1$, $S$ declines to zero. Thus, if $S_{max}>S_{cr}$, 
there are critical values of $x$
\begin{equation}
x_{1,2}=\frac{1}{2}\left(1\mp\sqrt{1-\frac{S_{cr}}{S_{max}}}\right),
\label{roots}
\end{equation} 
such that for $x_1<x<x_2$ lensing is caustic while for 
$x<x_1$ and $x>x_2$ lensing is in the noncaustic regime. If
$S_{max}<S_{cr}$ lensing is always noncaustic.

For the observer to see a noncaustic lensing event with a magnification larger
than some threshold magnification $A_t$, the lens has to pass
near the source star with a sufficiently
 small  (unlensed) impact parameter $b_0$.
In other words, magnification $M>A_t$ if 
$b_0 < b_{0 t}(A_t)$, where $b_{0 t}(A_t)$ is
given by 
\begin{equation}
M(b_{0 t}(A_t))=A_t.
\label{b_0}
\end{equation} 

As the cloud moves through the sky, all
the source stars lying in a strip on the sky along the lens trajectory with
 angular width $2\delta$ given by 
\begin{equation}
\delta(A_t,x) = b_{0 t}(A_t)/D_{OL},
\end{equation}
 are amplified with 
magnification $M > A_t$.

Let ${\bf v}_S$, ${\bf v}_L$, and ${\bf v}_O$ be the 
 transverse velocities (i.e. velocities 
perpendicular to the line from observer to source star)
of the source star, the  lens,
 and the  observer. Then the relative transverse velocity ${\bf v}_{\perp}$ 
of the gaseous
lens and the source star as seen by the observer is just
\begin{equation}
{\bf v}_{\perp}={\bf v}_L-\left[x {\bf v}_S + (1-x) {\bf v}_O\right]
\label{vels}
\end{equation}  
(Han \& Gould 1996).

It is important to distinguish  events with different
durations. Any real gravitational microlensing experiment has some 
finite probability $<1$ of detecting a lensing event which occurred during
the monitoring campaign and this  detection efficiency function $\phi$
depends strongly upon the timescale of the observed event (Alcock 1997).
We define the timescale $\tau$ of the lensing event as the 
time which the light curve spends above a threshold magnification $A_t$:
\begin{equation}
\tau=\frac{2 D_{OL}\sqrt{\delta^2-\varphi^2}}{|{\bf v}_{\perp}|},
\label{timescale}
\end{equation}
where $\varphi$ is the angular impact parameter of the lens's trajectory 
relative to the source star; $\varphi < \delta$
(otherwise $M < A_t$).

In calculations of the rate of lensing we must take into account 
the fact that 
lensing clouds and source stars have some distribution in velocity space;
we average the lensing rate over these distributions to obtain the
true expected rate.
The general
formula for the  event rate per lensing cloud is
\begin{equation}
d\dot N=2 \Sigma_S  \int\limits^{\delta}_0
d\varphi
\int d{\bf v}_L\int d{\bf v}_S f_L({\bf v}_L) f_S({\bf v}_S)
\frac{|{\bf v}_{\perp}|}{ D_{OL}}
\phi_{\tau}(\tau(\varphi,|{\bf v}_{\perp}|)),
\label{rate-raw}
\end{equation}
where $f_L$ and $f_S$ are the velocity distribution functions
for lensing clouds and source stars, respectively,
$\Sigma_S$ is the 
number of source stars per unit solid angle on the sky, and 
$\phi_{\tau}$ is the detection efficiency function for a lensing event 
with timescale $\tau$.

We assume for simplicity
 isotropic Maxwellian distribution functions for the
velocities of both lenses and source stars with dispersions
$\sigma_L$ and $\sigma_S$ respectively, taking into account 
the fact that the object containing stars is moving
in general, i.e. there is some offset velocity 
${\bf v}_c$ in the latter distribution:
\begin{equation}
f_L({\bf v}_L)=\frac{1}{\pi^{3/2}\sigma_L^{3/2}}
e^{-{\bf v}_L^2/\sigma_L^2},
\label{dis-l}
\end{equation}
and
\begin{equation}
f_S({\bf v}_S)=\frac{1}{\pi^{3/2}\sigma_S^{3/2}}
e^{-({\bf v}_S-{\bf v}_c)^2/\sigma_S^2}.
\label{dis-s}
\end{equation}
Our calculations are not very sensitive to this 
assumption in the sense that the order of magnitude result 
does not depend strongly upon the exact shape of the velocity 
distribution. The important assumption of isotropy of the 
distribution function permits
analytical simplifications.

Consider a coordinate system with the $z$-axis lying along the 
line of sight from observer to the source, $x$-axis perpendicular to the 
line of sight so that observer's velocity
${\bf v}_O$ lies in the $xz$-plane, and $y$-axis perpendicular to those two,
so that $v_{Oy}=0$. 
Taking distributions (\ref{dis-l}) and 
(\ref{dis-s}) it is shown in Appendix \ref{A1} that 
the total event rate per source is
\begin{eqnarray}
\dot N_{tot}
=4 D_{S}^2\int\limits^1_0 dx \frac{x n_L({\bf x})
\delta(x,A_t) v_{ch}^3}
{\sigma_L^2+x^2\sigma_S^2}e^{-C}
\int\limits^{\infty}_0 
u^2\kappa(u v_{ch})e^{-A u^2}I_0\left(B u\right)du,
\label{rate-tot}
\end{eqnarray}
where $I_0$ is the modified Bessel function of  order zero,
$n_L({\bf x})$ is the number density of lensing
clouds,
$v_{ch}(x)$ is some characteristic velocity (the final result does not
 depend upon the choice of this velocity), function $\kappa$
is defined as
\begin{equation}
\kappa(u)=
\int\limits^1_0
\phi_{\tau}
\left(\frac{2\delta(A_t,x) x D_{S}}{u}\sqrt{1-s^2}\right)ds,
\end{equation}
and  $A(x),B(x)$ and $C(x)$ are
given by
\begin{eqnarray}
A=\frac{v_{ch}^2}{\sigma_L^2+x^2\sigma_S^2},~~~~~~~
B=\frac{2v_{ch}}{\sigma_L^2+x^2\sigma_S^2}
\sqrt{x^2 v_{c y}^2 + (x v_{c x}+(1-x)v_{Ox})^2},\nonumber\\
C=\frac{x^2 v_{c y}^2 + (x v_{c x}+(1-x)v_{Ox})^2}{\sigma_L^2+x^2\sigma_S^2}.
\label{constants}
\end{eqnarray}
This formula for the 
lensing rate is quite general. For example, it is directly 
applicable to the case of gravitational microlensing [with corresponding
$\delta(A_t,x)$], in which case it  is identical to the formula 
for the event rate obtained by Griest (1991).

It is interesting to know the timescale distribution of the event rate
to have some idea of where to look for the events produced 
by the gaseous lensing.  The timescale distribution
 is given by (see Appendix \ref{A1})
\begin{eqnarray}
\frac{d\dot N_{tot}}{d\tau}
=32 D_{S}\frac{\phi_\tau(\tau)}{\tau^4}
\int\limits^1_0 dx \frac{n_L({\bf x})
b_{0 t}^4(x,A_t)}
{\sigma_L^2+x^2\sigma_S^2}e^{-C}
\int\limits^1_0 
\frac{u^4}{\sqrt{1-u^2}}e^{-A u^2}I_0\left(B u\right)du,
\label{dist}
\end{eqnarray}
where $b_{0 t}(x,A_t)$ is defined in (\ref{b_0}) and $v_{ch}$, 
entering the definitions of $A$, $B$, and $C$ in (\ref{constants}), 
is given by  
$v_{ch}=2 b_{0 t}(x,A_t)/\tau$ [compare with a similar formula by Griest
(1991) for the case of gravitational microlensing].

Closely related to the timescale distribution $d\dot N_{tot}/d\tau$
is its integral from $\tau_{min}$ 
to infinity, that is the frequency of events with 
timescales exceeding $\tau_{min}$, 
\begin{eqnarray}
\dot N_{tot}(\tau>\tau_{min})=
\int\limits_{\tau_{min}}^{\infty}\frac{d\dot N_{tot}}{d\tau}d\tau \nonumber\\
= 32 \frac{D_{S}}{\tau_{min}^3} \int\limits^1_0 dx \frac{n_L({\bf x})
b_{0 t}^4(x,A_t)}
{(\sigma_L^2+x^2\sigma_S^2)}e^{-C}
\int\limits^1_0 
u^2 \tilde \kappa(u) e^{-A u^2}I_0\left(B u\right)du,
\label{rate-t_min}
\end{eqnarray} 
where function $\tilde \kappa$ is now
\begin{eqnarray}
\tilde \kappa(u)=\int\limits^{\sqrt{1-u^2}}_0 \phi_{\tau}
\left(\frac{\tau_{min}}{u}\sqrt{1-s^2}\right) ds.
\end{eqnarray}

If one wants to know the true event rate, for $100\%$ efficiency,
one need only set $\phi_\tau=1$ in formulae (\ref{dist}) and
(\ref{rate-t_min}), as we will do in our consideration of
the event rate distributions.

In the next section we show typical plots of $\dot N_{tot}(\tau>\tau_{min})$
for some clouds  characterized by polytropes with $n=1.5$ and $n=4.5$.

\section{Results}\label{results}

In this section we consider 
 the lensing of  stars located in the LMC
and
compare with the results obtained by the MACHO collaboration, which 
has monitored $11.9$ million LMC stars for $5.7$ yrs (Alcock et al. 2000).

\subsection{Parameters of the event rate calculations}\label{param}

The LMC has a distance  $D_{LMC} \approx 55$ kpc 
from the Sun, and its position 
in galactic coordinates is
$b=-32.8^{\circ}$ and $l=281^{\circ}$.
 Jones et al. (1994) finds the motion of the LMC to be 
\begin{eqnarray}
v_l= - 16 \pm 60 ~{\rm km}~{\rm s}^{-1}\nonumber\\
v_b= 328 \pm 60~ {\rm km}~{\rm s}^{-1}\\
v_{rad}= 250 \pm 10~ {\rm km}~{\rm s}^{-1}.\nonumber
\end{eqnarray}
We use these proper motion measurements for obtaining our event rates. 
The velocity dispersion of the lensing clouds was assumed to be $\sigma_L=220~
{\rm km}~{\rm s}^{-1}$ and that of the stars in the LMC was taken
to be $\sigma_S=10~ {\rm km}~{\rm s}^{-1}$.

We performed our calculations both for the case of the true event rate
(for $100 \%$ detection efficiency function) and also 
for the the case of the MACHO
detection efficiency function which is given in Alcock et al (1997).
In the first case we assume a threshold magnification 
$A_t=1.1$ and in the second we take $A_t=1.35$, which is the current 
threshold magnification of the MACHO experiment (Alcock et al. 2000). 
Estimation of  the MACHO detection efficiency function $\phi_{\tau}$
is problematic because 
it is known in terms of the  duration of a gravitational 
microlensing event, which differs significantly  from our 
definition (\ref{timescale}) for
the gaseous lensing timescale. The method by which we 
estimate $\phi_{\tau}$ is described in Appendix \ref{A2}, and our
adopted $\phi_{\tau}$ is shown on Figures \ref{t_d_1_5} and \ref{t_d_4_5}.

We model the lensing clouds by self-gravitating 
H$_2$-He (nonrotating) polytropes  of radius $R_{cl}$.
The polytropic index $n$ ($T\propto\rho^{1/n}$) 
must be in the range $1.5\le n<5$;
for $n<1.5$ the cloud would be convectively unstable, while for
$n\ge5$ the central density is infinite.
We do not expect $T(r)$ and
$\rho(r)$ to be accurately described by a polytropic model,
but a slight rise in temperature toward the interior may be reasonable
since the interior will be heated by high energy cosmic rays,
with the few cooling lines [e.g., H$_2$ 0-0S(0) 28.28$\micron$, or
HD 0-0R(1) 112$\micron$]
very optically thick.
Furthermore, polytropes have $T(R_{cl})=0$, so the density structure
near the surface is unphysical.
Table 1 of Draine (1998) gives various properties 
for H$_2$-He polytropes.
In our current calculations we consider gaseous clouds with polytropic 
indices $n=1.5$ and $n=4.5$, to compare extremes of behavior\footnote{
We assume that it is  the density-radius relation of the clouds which is 
determined by the polytropic law, not the behavior of the gas
on  dynamical
 timescales (otherwise the $n=4.5$ polytrope would be 
dynamically unstable.)
}.

Following Widrow \& Dubinski (1998)
we adopt a Navarro, Frenk, \& White (1996) model of a  spherical
halo, composed of gaseous 
clouds:
\begin{equation}
n_{L}({\bf r})=f_{cl}
\frac{M_{\rm MW}}{4\pi M_{cl}}\frac{1}{r(r+a_s)^2}
\label{halo}
\end{equation} 
where $r$ is the distance between the galactic center and the point
of interest,
$a_s=15.9$ kpc is the core radius,
and  $0 \le f_{cl} <1$ is the fraction of the total mass of the 
Milky Way
$M_{\rm MW}=6.5 \times 10^{11}~{\rm M}_{\odot}$ 
contributed by the population of  self-gravitating 
H$_2$ clouds.

\begin{center}
\begin{deluxetable}{ c c c c c c c}
\tablecolumns{6}
\tablewidth{0pc}
\tablecaption{Models used for timescale distribution.  
\label{table}}
\tablehead{
\colhead{model}&
\colhead{n}&
\colhead{$M_{cl}\tablenotemark{1} $}&
\colhead{$R_{cl}\tablenotemark{1} $}&
\colhead{$ T_c\tablenotemark{2} $}&
\colhead{$2GM_{cl}m_H/3 k_B R_{cl}\tablenotemark{3} $}&
\colhead{$S_{max}$}
	}
\startdata
$A$  & $1.5$ & $10^{-2}$ & $6.6$ & $200$ & $109$ & $2.6\times 10^{-3}$\\
\hline
$B$  & $1.5$ & $10^{-2}$ & $2.6$ & $507$ & $276$ & $0.106$ \\
\hline
$C$  & $1.5$ & $10^{-2}$ & $1.1$ & $1200$ & $652$ &  $3.3$\\
\hline
$D$  & $1.5$ & $10^{-3}$ & $3.7$ & $36$ & $19$ & $2.6\times 10^{-3}$ \\ 
\hline
$E$  & $1.5$ & $10^{-3}$ & $1.5$ & $88$ & $48$ & $0.096$ \\
\hline
$F$  & $1.5$ & $10^{-3}$ & $0.6$ & $220$ & $119$ & $3.8$ \\
\hline
$G$  & $1.5$ & $10^{-4}$ & $0.8$ & $17$ & $9$ & $0.12$ \\
\hline
$H$  & $1.5$ & $10^{-4}$ & $0.4$ & $33$ & $18$ & $1.9$ \\
\hline
$I$  & $4.5$ & $10^{-2}$ & $64$ & $13$ & $11$ & $2.9\times 10^{-7}$ \\
\hline
$J$  & $4.5$ & $10^{-2}$ & $8.8$ & $93$ & $81$ &  $8.1\times 10^{-4}$\\
\hline
$K$  & $4.5$ & $10^{-2}$ & $1.3$ & $628$ & $552$ & $1.7$ \\
\hline
$L$  & $4.5$ & $10^{-3}$ & $5$ & $16$ & $14$ & $7.8\times 10^{-4}$ \\
\hline
$M$  & $4.5$ & $10^{-3}$ & $0.7$ & $117$ & $102$ & $2.0$ \\
\hline
$N$  & $4.5$ & $10^{-4}$ & $0.4$ & $20$ & $18$ & $1.9$ \\
\enddata
\tablenotetext{1}{$M_{cl}$ is in $M_{\odot}$, $R_{cl}$ is in AU. }
\tablenotetext{2}{Cloud central temperature, in Kelvins}
\tablenotetext{3}{Temperature at which the thermal energy 
of a hydrogen atom at the
cloud surface equals its gravitational binding energy, in Kelvins}
\end{deluxetable}
\end{center}

\subsection{Timescale distributions}

\begin{figure}
\vspace{18.cm}
\includegraphics{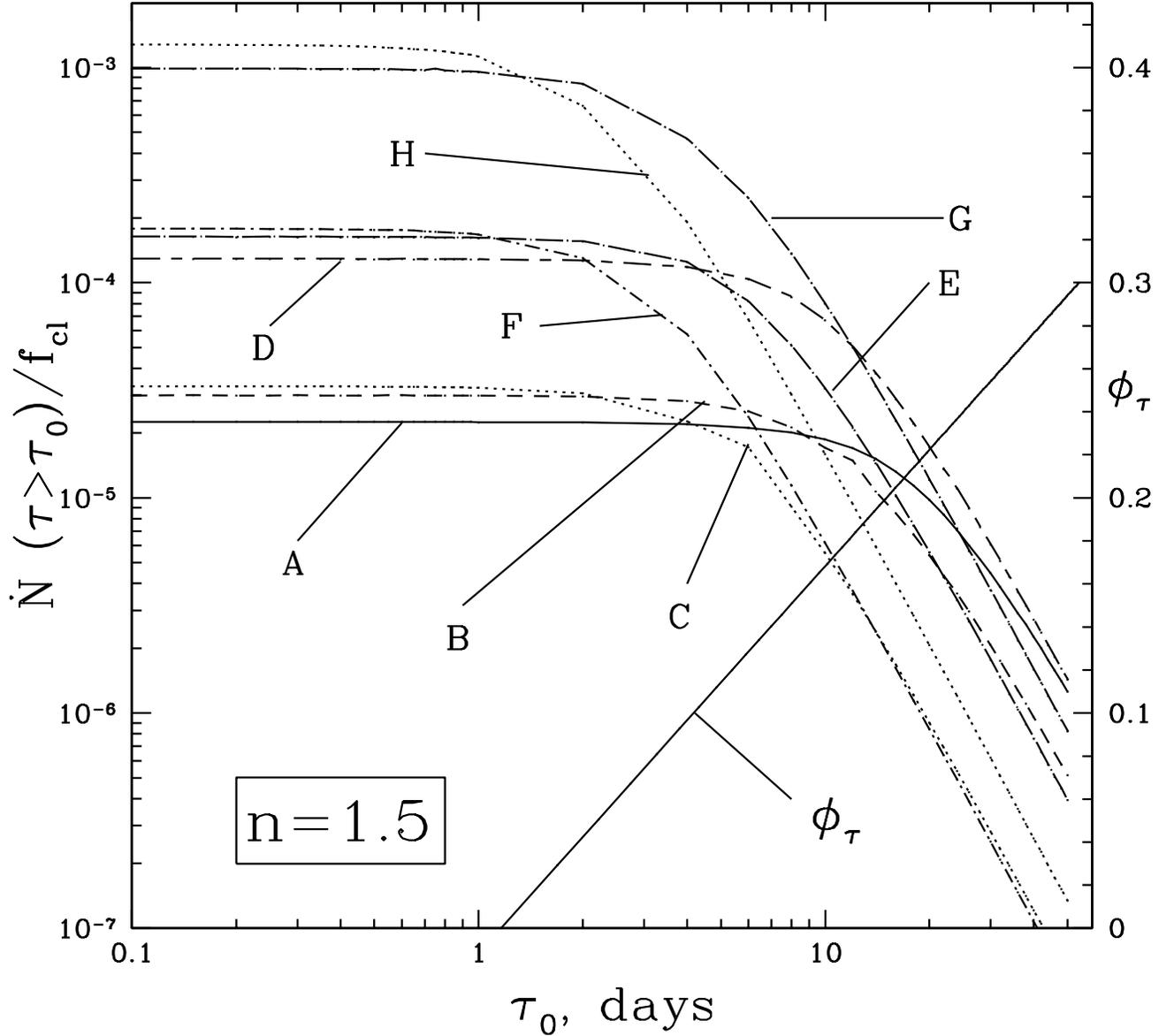}
\caption{
Distribution of $\tau$ for lensing by selected  $n=1.5$ polytrope
clouds. The timescale $\tau$ is defined to be the time during which the 
magnification $A>A_t=1.1$.
The rate of events with timescales larger than 
$\tau_0$ is shown, 
for
detection efficiency function $\phi_{\tau}=1$.
Timescales of a few days are typical; for model F, for example, $50 \%$
of events have $\tau<3$ days. The  solid line labelled $\phi_\tau$
shows the detection efficiency function used 
in the MACHO event rate calculations in Figures \ref{n_1_5_MACHO}
and \ref{n_4_5_MACHO}. 
}
\label{t_d_1_5}
\end{figure}

\begin{figure}
\vspace{18.cm}
\includegraphics{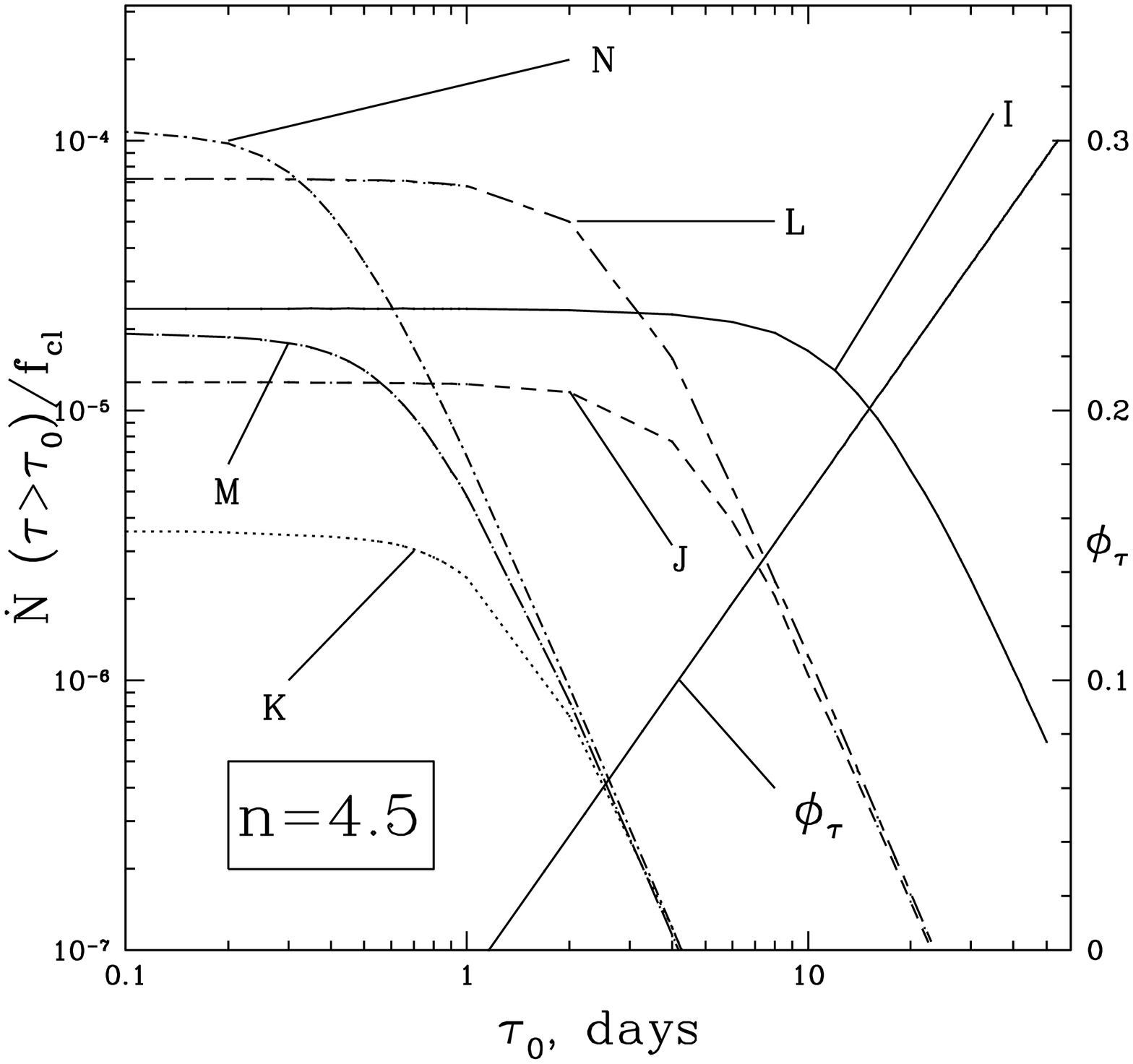}
\caption{
Same as Figure \ref{t_d_1_5} but for $n=4.5$.
}
\label{t_d_4_5}
\end{figure}

On Figures \ref{t_d_1_5} and \ref{t_d_4_5} we plot the dependence 
of the integrated timescale distribution $\dot N_{tot}(\tau>\tau_{min})$,
as a function of $\tau_{min}$, for  some representative values 
of the cloud mass $M_{cl}$ and radius $R_{cl}$. 
The cloud parameters are given 
in  Table \ref{table} and we assume $\phi_\tau=1$.

All timescale distribution curves exhibit a plateau
for small timescales and then a  rapid falloff at larger 
event durations.
From these plots we see that the characteristic timescale of the
lensing events is quite small, 
varying from several days 
to several tens of days. This is as
expected, because cloud sizes are small, $\sim 10$ AU, 
sufficient
amplification occurs only when $b_0 < (0.1 - 0.3) R_{cl}$,
and for typical transverse velocity 
$\approx 200$ km ${\rm s}^{-1}$ we get just 
this range of event durations. 

The time distribution seen by a real experiment would be modified
by the detection efficiency, 
which may be small for short timescale events.
In addition, events lasting $\ge 50$ days, which occur when the transverse 
velocity of the lens 
is quite small, will be affected by changes in the velocity of the Earth 
over this period (the so-called ``parallax effect'' (Paczy\'nski 1996; 
Gould \& Andronov 1999)).

\subsection{Results for the event rate}\label{ev-rate}

Calculation of the event rate is quite straightforward in the 
case of noncaustic lensing because in this case foreground star
crossing the cloud produces magnifications $A>A_t$ in one 
continuous span (see Figure \ref{lightcurve} for the cases
$S<S_{cr}$). In the caustic regime  the situation
is more complicated, since for clouds located at the middistance 
between observer and foreground star these magnifications $A>A_t$
will be not only in the region of the closest approach but 
also at the caustic spikes (see Figure \ref{lightcurve} for the cases
$S>S_{cr}$) and we want to account for this.

\begin{figure}
\vspace{13.cm}
\includegraphics{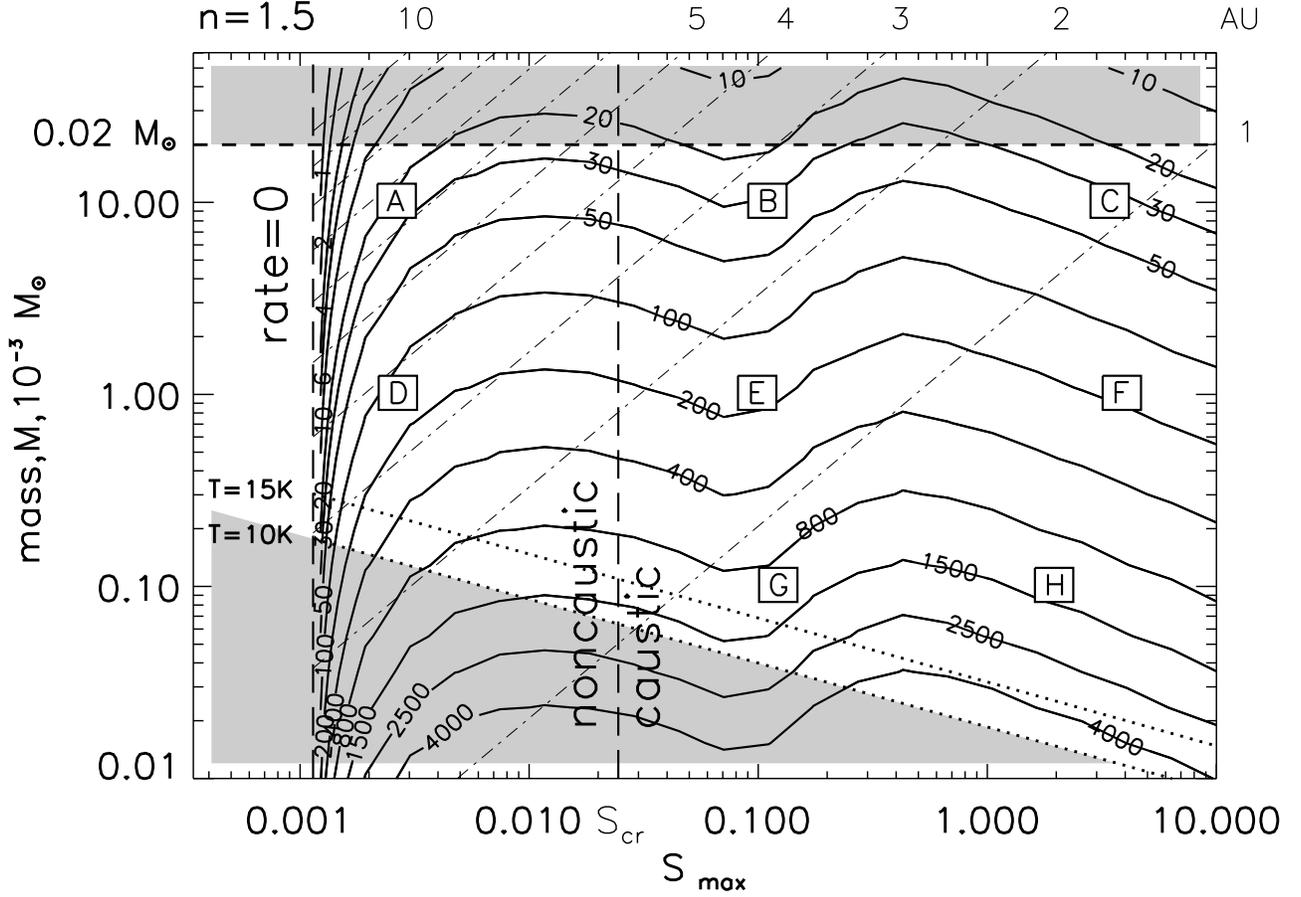}
\caption{
Contour plot of the rate of events with peak magnification $A>A_t=1.1$ and
any $\tau$ 
produced by gaseous lensing 
as function of $M_{cl}$ and $S_{max}
\equiv 3 \alpha M_{cl}D_{LMC}/16\pi R_{cl}^4$.
Each cloud is assumed to be an $n=1.5$ polytrope.
Thick solid lines represent levels of constant event rate labelled 
in units of $10^{-6} ~{\rm events}~{\rm yr}^{-1}$ per source, for $f_{cl}=1$.
Dash-dotted lines are the lines of constant $R_{cl}$ in these coordinates,
labelled by $R_{cl}$ in AU.
To the left from the thick dashed line ``rate $= 0$'' 
the clouds are unable to produce magnification $A_{max}>A_t=1.1$
 (see equation (\ref{S_l})). 
Dashed vertical line at $S_{max}=S_{cr}$ 
separates caustic and noncaustic regimes.
Dotted lines show the lower limit $M_{cl}/R_{cl}$
so that a molecule at the surface will be bound for an atmospheric
temperature
 of $10$ K or $15$ K. 
Dashed line at $M=0.02$
${\rm M}_{\odot}$ is the upper limit on $M_{cl}$ suggested by
 Wardle \& Walker (1999).
The unshaded region is allowed.
Squares with letters show the positions of the sample cloud
models from  Table \ref{table}.
Note the very high event rate for almost all cloud models
 (to the right from line ``rate $= 0$'').
}
\label{n_1_5}
\end{figure}

We adopted a simple approach for the caustic regime. One can see from 
Figure \ref{lightcurve} (upper panel) that for a given 
polytropic index there exists
a specific value of $S_\star>S_{cr}$ such that the inner minimum of the 
lightcurve (between the central part and one of the caustic spikes)
has magnification $A_t$. For example, in the case $n=1.5$  
$S_\star=0.112$ for $A_t=1.1$ or $S_\star=0.101$ for $A_t=1.35$.
 It is obvious that when $S_{cr}<S<S_\star$
the picture  is similar to
the noncaustic case since magnifications $A>A_t$ are reached in one 
continuous span. If $S>S_\star$ there are also caustic spikes
producing events but one can easily see that the duration of events 
caused by these spikes  decreases rapidly with growing $S$  (Draine 1998)
and we simply ignore
them in our calculations. That is we assume for $S>S_\star$ that
 $A>A_t$ only in the central part of the lightcurve. This gives us 
only  a lower limit on the event rate but a useful one since the 
number of events produced by caustics is 
relatively small and the caustic spikes are of short duration.

\begin{figure}
\vspace{13.cm}
\includegraphics{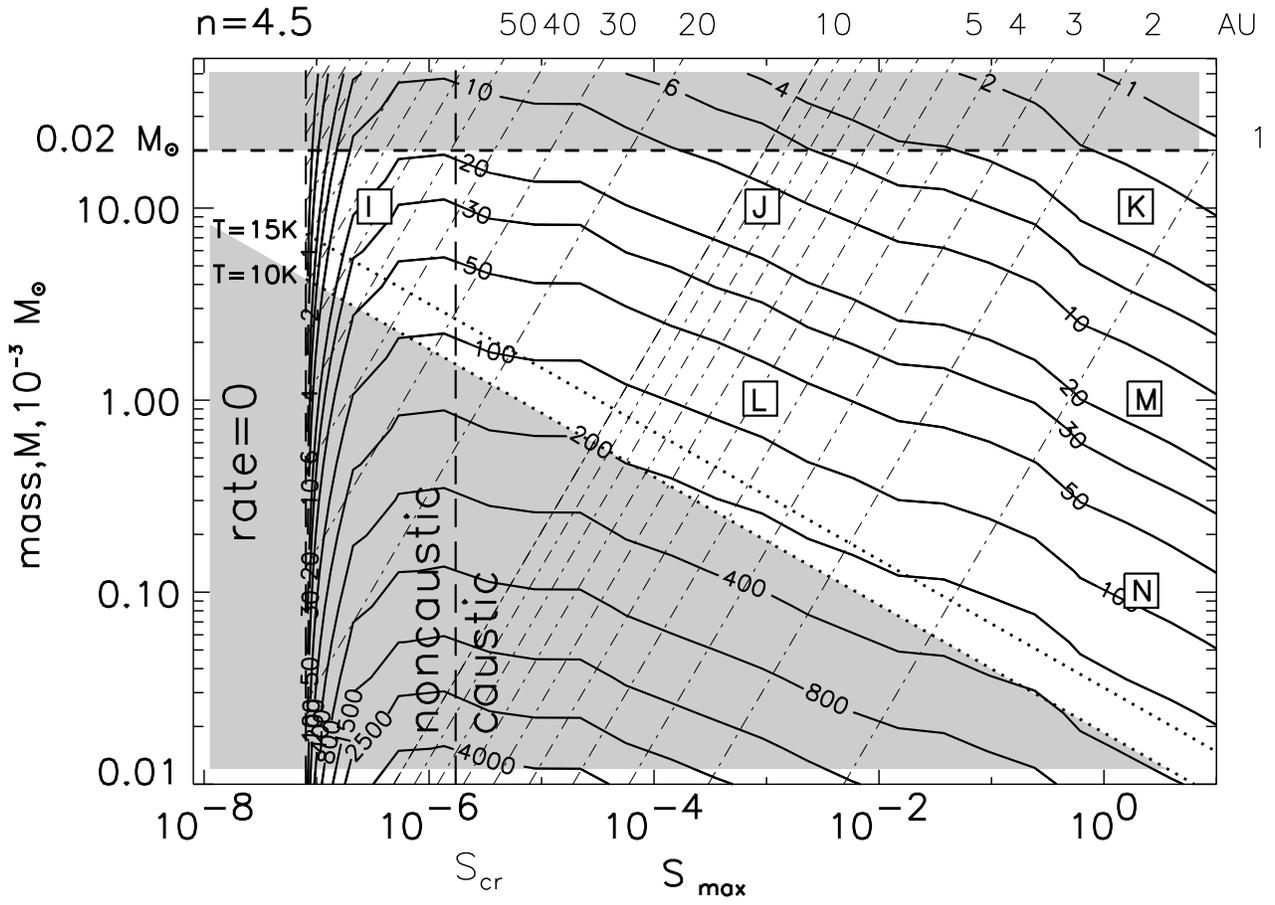}
\caption{
Same as Figure \ref{n_1_5} but for $n=4.5$.
}
\label{n_4_5}
\end{figure}

In Figures \ref{n_1_5} and \ref{n_4_5} we show the true
event rate (i.e. for efficiency $\phi_{\tau}=1$) due to 
gaseous lensing for various cloud parameters
and for two polytropic indices of the clouds: $n=1.5$ and
and $n=4.5$. 
We characterize the clouds by mass $M_{cl}$ and $S_{max}$, the 
``strength'' parameter for a cloud located midway between the Earth 
and the LMC: 
$S_{max}=\alpha\langle\rho\rangle D_{LMC}/4R_{cl}$,
where $D_{LMC}=55$ kpc. For each value of
$S_{max}$ and $M_{cl}$  the  cloud radius $R_{cl}$
is obtained from equation (\ref{eq:Sdef}):
\begin{eqnarray}
R_{cl}=\left(\frac{3\alpha M_{cl} D_{LMC}}{16 \pi S_{max}}\right)^{1/4}
=1.25\times 10^{13}~ {\rm cm}~ \left(\frac{M_{cl}}{10^{-3}M_{\odot}}
\right)^{1/4}\frac{1}{S_{max}^{1/4}}.
\end{eqnarray}
 Lines of constant cloud radius 
are plotted on each of Figures \ref{n_1_5}-\ref{n_4_5}
as dot-dashed lines.
We consider here events with all timescales and magnifications 
larger than $A_t=1.1$.

Our adopted  bound on $M_{cl}$ comes from two considerations. First,
clouds which are 
too massive  would be similar to the objects 
known to be unstable to collapse to form  low
mass stars. 
Second, Wardle \& Walker (1999) obtained an upper bound on 
$M_{cl}<10^{-1.7} M_{\odot}$ by considering a
 cooling mechanism which could stably radiate away the heat deposited
by cosmic rays.

We restricted the parameter space of $S_{max}$
and $M_{cl}$ because of the following reasons.
The upper bound in $S_{max}$ is just $10$ because this 
 corresponds to a cloud with $M_{cl}=10^{-1.7} M_{\odot}$
and radius $R_{cl}=1$ AU. We do not consider smaller
 radii because extreme scattering events 
favor clouds with sizes of the order of several AU (Draine 1998).

The minimum value of  $S$ for a cloud to be detectable by light curve 
monitoring with a  
threshold magnification $A_t$ 
can be obtained from equation (\ref{M0}):
\begin{equation}
S_{det}=S_{cr}(1-1/\sqrt{A_{t}}),
\label{S_l}
\end{equation}
where $S_{cr}$ is given by equation (\ref{s-cr}). For $S_{max}>S_{det}$
the cloud population will produce a nonzero rate of events with
$A_{max}>A_t$.

Dotted lines show the restriction imposed by the requirement that the
gravitational binding energy of gas at the surface exceed the thermal 
energy for an estimated atmospheric temperature $T$ (Draine 1998):
\begin{equation}
M_{cl} > 6 \times 10^{-5}\left(\frac{T}{10 {\rm K}}\right)~
\left(\frac{R_{cl}}{{\rm AU}}\right)~M_{\odot}.
\end{equation}
Two  boundaries are shown, for atmospheric
temperatures $T=10$ K and $T=15$ K.  
The shaded region below these curves is prohibited.

The thick solid lines 
are contours of constant event rate.
The lensing rate is very high for all the cloud models appreciably to
the right of the ``rate $= 0$'' line.

\begin{figure}
\vspace{13.cm}
\includegraphics{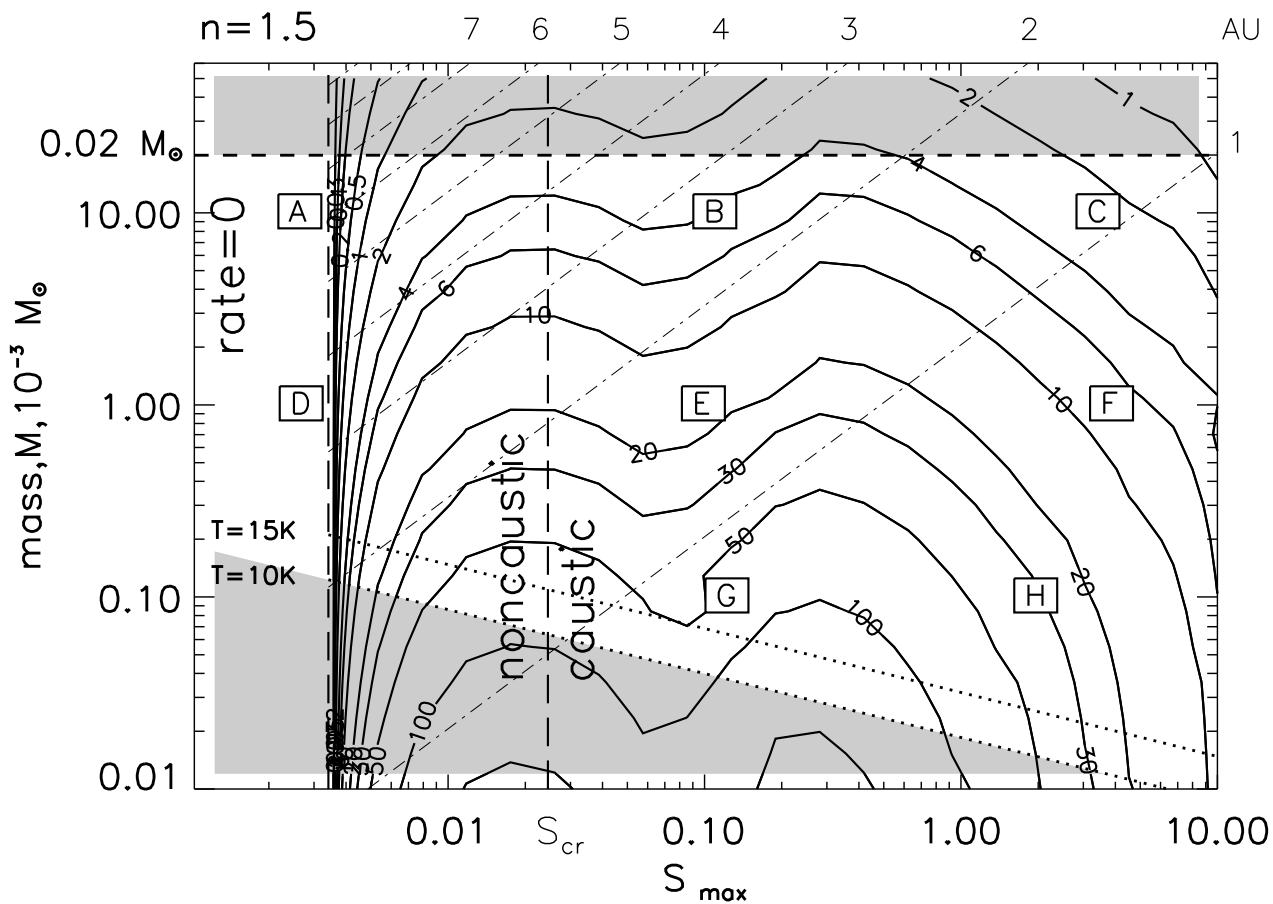}
\caption{
Contour plot of the event rate of gaseous lensing 
for various parameters of the lensing clouds,
for the  MACHO detection efficiency function.
Polytrope $n=1.5$ is assumed and threshold magnification 
$A_t=1.35$. All the symbols and lines are 
the same as in Figure \ref{n_1_5}. Note the substantial decrease of the 
event rate due MACHO's reduced sensitivity to short timescale events. 
}
\label{n_1_5_MACHO}
\end{figure}

Figures \ref{n_1_5_MACHO}  and \ref{n_4_5_MACHO} show the 
results for the lensing rate for polytropes with $n=1.5$
and $n=4.5$ for a detection efficiency function approximately that of
 the 
MACHO experiment. All the notation is the same as in the case 
of $\phi_{\tau}=1$. The ``rate $= 0$'' boundary has moved (relative
to Figures \ref{n_1_5} and \ref{n_4_5}) because we now require 
a threshold magnification $A_t=1.35$.
Small glitches on the contours are artifacts of the numerical procedure.
We consider here the events with all timescales filtered through 
the detection efficiency function discussed in Appendix \ref{A2}.

The rate which would be seen by the MACHO experiment is 
significantly smaller than the rate for $A_t=1.1$ and $\phi_{\tau}=1$,
 sometimes by two orders of magnitude.
This is primarily a consequence of the reduced sensitivity
to short timescale 
events of the MACHO detection efficiency function. 
Nevertheless
the predicted rate is still quite high and in the 
next subsection we  compare the predicted rate to that actually observed.

\subsection{Comparison with the MACHO results}

How shall we compare the results of the MACHO experiment with
 our simulated plot of what it would see if the dark matter is composed
of molecular clouds?
First of all, we need to estimate
what fraction of the events actually observed by MACHO could
be attributed to
 gaseous lensing. This can be done by noticing that the 
typical timescale of gaseous lensing events is very small, while the 
shortest timescale  event reported by MACHO
has $\hat t=34$ days. For 
the types of timescale distribution seen in Figures \ref{t_d_1_5}
and \ref{t_d_4_5} it is extremely unlikely
that there are more than $1$ or $2$ gaseous lensing 
events among those observed by MACHO, since otherwise a large number
of short timescale events (with durations less than $10$ days) would be seen,
in conflict with the distribution of timescales observed by the
MACHO and EROS (Alcock et al. 2000;
Lasserre et al. 2000)
collaborations. It  would also 
be completely inconsistent with the EROS and MACHO 
combined limits on the rate of very short timescale events, from
$15$ minutes to several days, for which they claim 
that the analysis of two years of data on $8.6$ million stars found 
no short-duration ``spike'' events (Alcock et al. 1998), implying an 
upper limit $\dot N \le 10^{-7}$ events yr$^{-1}$.

\begin{figure}
\vspace{13.cm}
\includegraphics{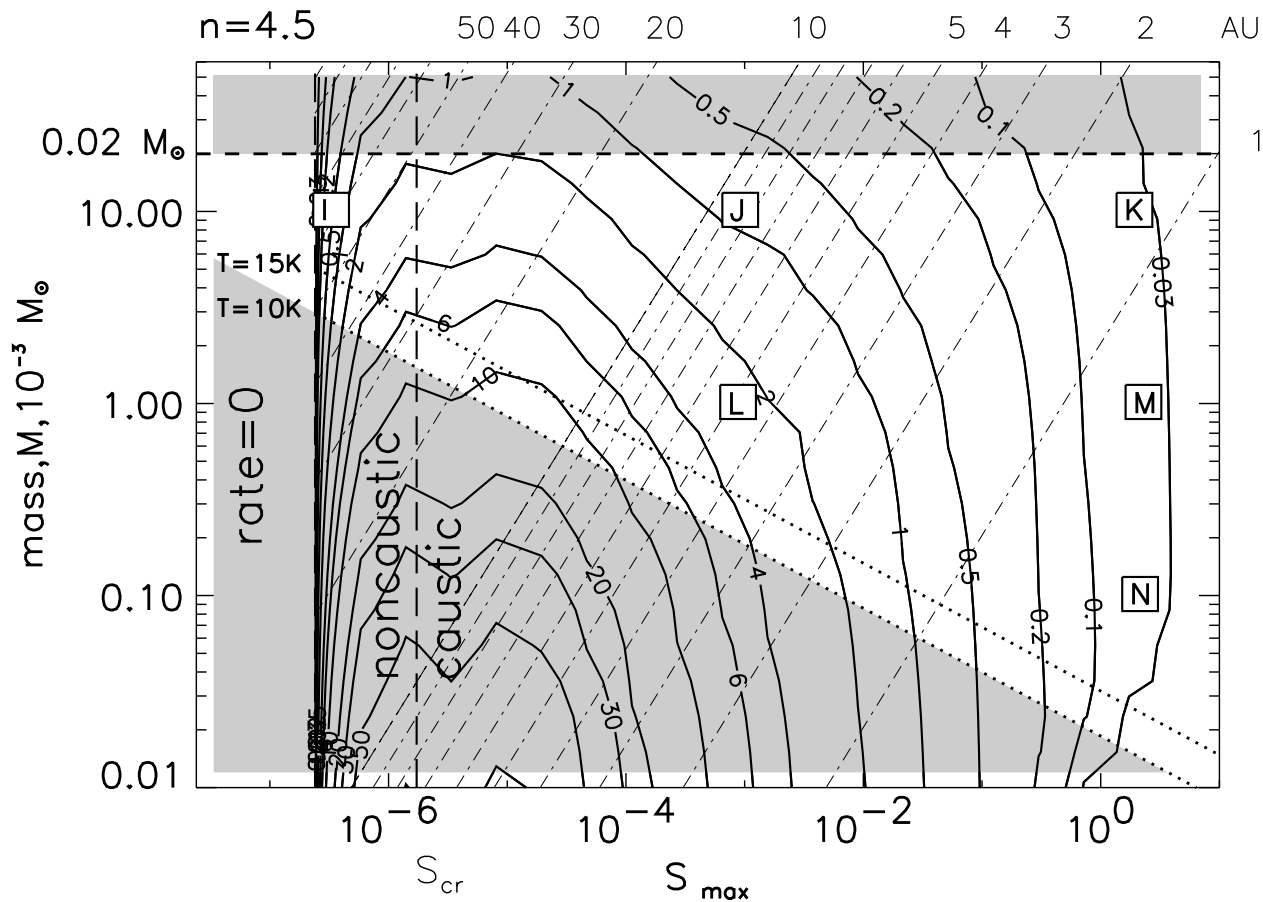}
\caption{
Same as Figure \ref{n_1_5_MACHO} but for $n=4.5$. 
Notice another allowed region 
 at small cloud radii, bounded by contour corresponding
to the event rate $3\times 10^{-8}
 ~{\rm events}~{\rm yr}^{-1}$ and line $M=10^{-1.7}~ M_{\odot}$.
}
\label{n_4_5_MACHO}
\end{figure}

The MACHO collaboration observed $11.9$ million
stars in the LMC for $5.7$ years. This means that the 
rate which could possibly
be due to  
 gaseous lensing events can be at most 
$(0.01-0.03) 
\times 10^{-6} ~{\rm events}~{\rm yr}^{-1}$. It is clear from 
our simulated observations of gaseous lensing on Figures
\ref{n_1_5_MACHO} and \ref{n_4_5_MACHO} that if $f_{cl}\ge 0.1$
the contour
describing such a rate will lie very close to the line 
$S_{max}=S_{det}$, where the event rate drops to $0$.
Any cloud model with $S_{max}<S_{det}$ will of course be 
undetectable and thus be allowed. Models with  
$S_{max}> 1.1 S_{det}$
 are prohibited, because otherwise MACHO would have seen an enormous number
of gaseous lensing events.

\begin{figure}
\vspace{14.cm}
\includegraphics{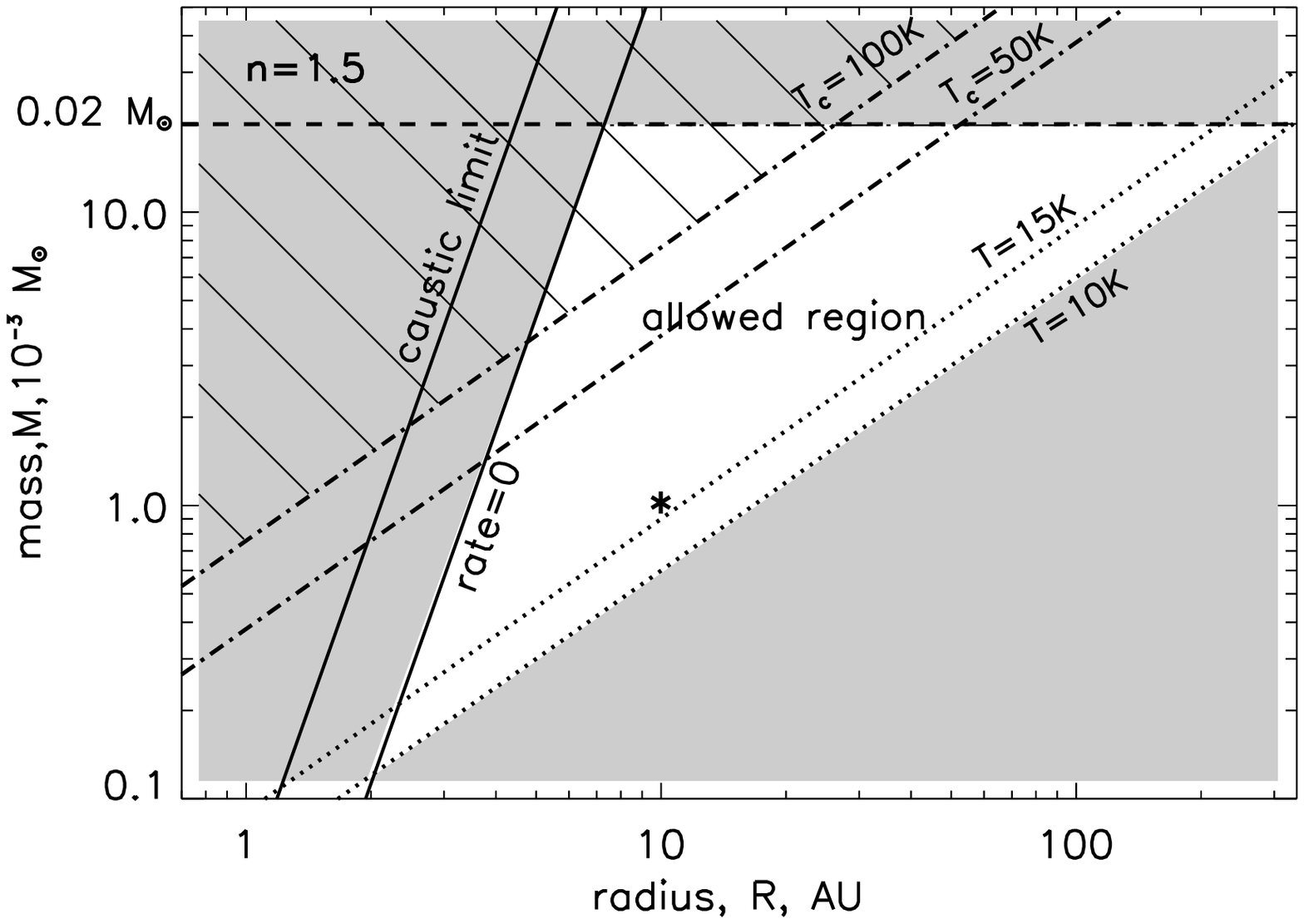}
\includegraphics{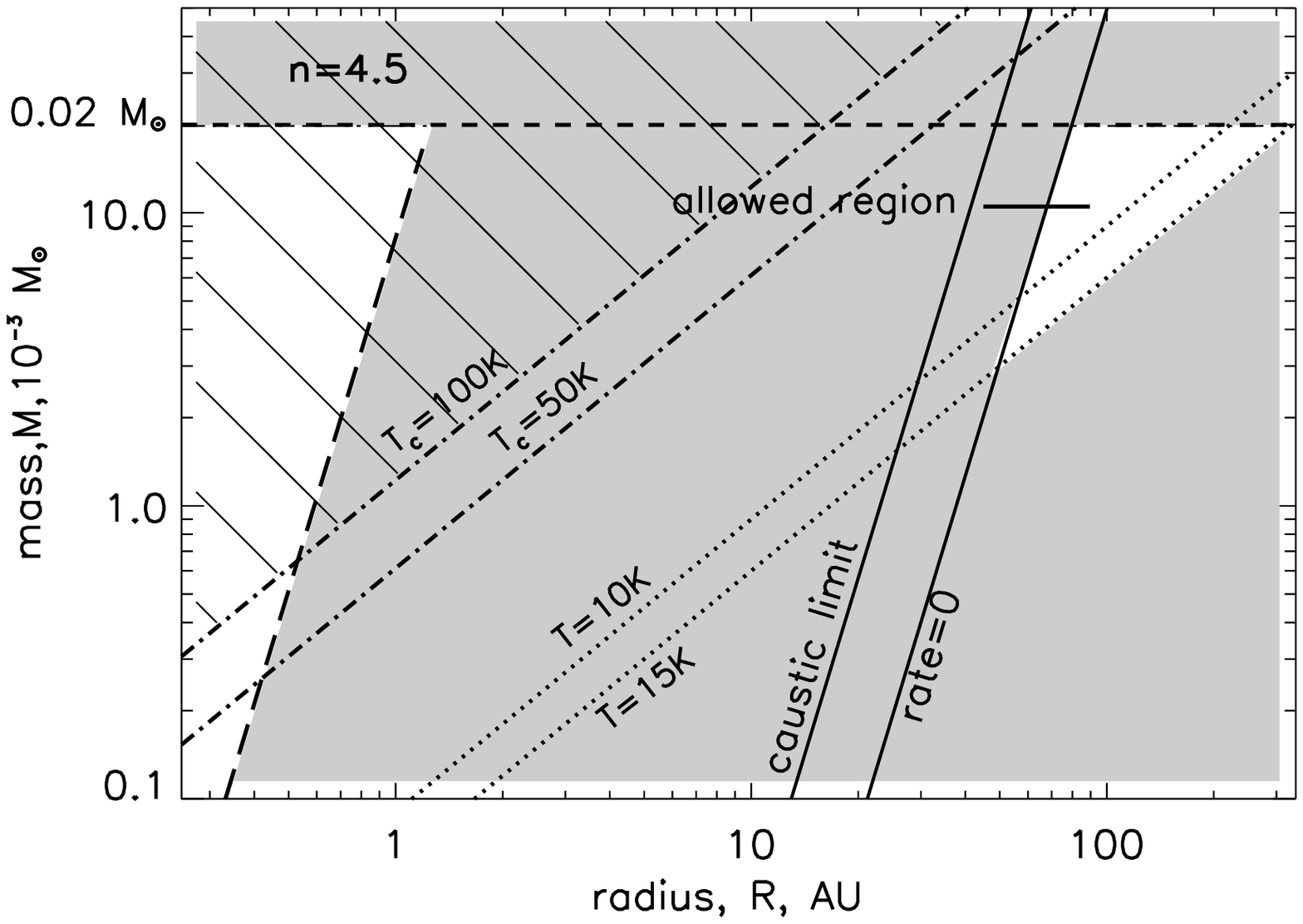}
\caption{
Exclusion plot for cloud radius $R_{cl}$ and mass $M_{cl}$,
showing region of allowed parameters (unshaded) and prohibited
(shaded regions) for clouds with polytropic index  a) $n=1.5$ and 
b) $n=4.5$.
Dotted  lines represent limitation set by 
requirement that the atmosphere be gravitationally bound.
Dashed line shows mass restriction by 
Wardle \& Walker (1999). 
Region above the dash-dotted lines contains models with central temperature 
larger than $50$ K and $100$ K, respectively, which might exclude them
(line shading). 
Note that another allowed region 
appears for small cloud radii in the $n=4.5$ case, 
bounded by dashed line roughly corresponding
to the event rate $3\times 10^{-8}
 ~{\rm events}~{\rm yr}^{-1}$ and line $M=10^{-1.7}~ M_{\odot}$.
As discussed in the text, the $n=1.5 $
polytropic model is the most conservative assumption in the absence of 
detailed thermal models for these clouds. Star on the $n=1.5$ plot
shows the position of the cloud model with $R_{cl}=10$ AU and
$M_{cl}=10^{-3} M_{\odot}$.
}
\label{par_space}
\end{figure}

These considerations and the restrictions described in \S
\ref{ev-rate}
 are combined together to produce an exclusion plot (Figure \ref{par_space}). 
One can see
that an allowed region exists where clouds basically cannot be seen
by the MACHO experiment and are not prohibited by the maximum mass 
and ``thermal evaporation'' constraints. For larger $n$ 
(softer equation of state) the allowed region
shrinks.
It is likely that for $n \to 5$ this region 
disappears completely, though we did not run our calculations for 
$n$ larger than $4.5$.

Also in the case $n=4.5$ another allowed region appears for small 
cloud radii (see Figure \ref{n_4_5_MACHO}). It is bounded by the 
level contour corresponding to the rate $3\times 10^{-8}
 ~{\rm events}~{\rm yr}^{-1}$ and goes into the region of the small
cloud radii (see Figure \ref{par_space} b).

The allowed regions can be further narrowed by noting that clouds
are highly unlikely to have high central temperatures. 
Detailed studies of their thermal structure are not available,
but it seems reasonable to exclude models of clouds
with central temperatures $> (50-100)$ K. For $n=1.5$ this 
places another limit within the allowed region 
(see Figure \ref{par_space}a). In the case $n=4.5$ 
the allowed region for small radii
virtually disappears  because 
of this restriction (see Figure \ref{par_space}b).

In the absence of realistic models for the thermal structure of 
these clouds, we can only use $n=1.5$ polytropes as the most 
conservative model from the standpoint of gaseous lensing. 
For this case, we see that the allowed region is quite substantial,
and contains the model with $M_{cl}=10^{-3} M_{\odot}$ and
$R_{cl}=10$ AU which was favored by Draine (1998).

It is also possible to expand the allowed region by assuming that 
the clouds contribute only a small part
$f_{cl} \ll 1$ of the dark matter
(see \S \ref{param}),
but to get any significant effect we need to suppose that
this fraction is $<10^{-2}$ which would rule out these 
clouds as a main constituent of the dark matter, and eliminate
 them as the explanation for the extreme scattering events.

It is clear that our lensing restriction represented 
on Figure \ref{par_space} is 
sensitive to the adopted density profile of an individual cloud,
but it is relatively insensitive to changes
in the spatial and velocity distributions of the 
dark matter, since changes in these parameters are 
 unlikely to produce variations in the event rate of $\ge 2$
orders of magnitude, which would be needed to move the restriction
given by the observed lensing rate  
 significantly away from the line ``rate $= 0$''.

\section{Discussion}\label{disc}

Wardle \& Walker (1999) suggested that heating 
caused by cosmic rays in the cold molecular clouds 
can be balanced if particles of solid H$_2$ can form.
They argue that these particles cool the cloud by thermal continuum radiation
if the cloud mass lies in the range $10^{-7.5} - 10^{-1.7} M_{\odot}$.
Our calculations show
that the lensing event rate would grow to extremely high values
for masses below $10^{-4} M_{\odot}$ (see Figures
\ref{n_1_5_MACHO}  and \ref{n_4_5_MACHO} ).

Kalberla, Shchekinov, \& Dettmar (1999) proposed that the $\gamma$-ray
background emission from the Galactic halo seen in  EGRET
($E>100$ MeV photons) data can be produced by interaction of
high-energy cosmic rays with small dense clouds. They modeled
this effect with different types of spatial distribution of dark matter
in the form of these clouds and found that their best fit to the 
EGRET data for uniform density clouds occurred  for  
$M_{cl}/R_{cl}^2\approx 10^{-3}~ M_{\odot}/(6~ AU)^2$.
Their conclusions about the cloud parameters  are 
 model dependent, but the quoted values fall within the allowed 
range on Figure \ref{par_space} for $n=1.5$. 

Though it is impossible to observe the gravitational microlensing
by the molecular clouds in our Galaxy it becomes possible if one observes 
them in other galaxies. A search for such lenses in the 
Virgo cluster is now underway
(Tadros, Warren, \& Hewett 2000) and preliminary
results indicate that  galaxy halos are 
unlikely to primarily consist of object with masses smaller 
than $10^{-5}~M_{\odot}$. We confirm this restriction but also
place more rigorous constraints on the cloud properties since we can
reject many models with higher cloud masses.

Another possible approach would be  to look for 
periods of demagnification in the light curves of lensing events. 
For instance, the OGLE-II survey (Udalski et al. 1997) 
has collected a large sample 
of light curves for stars in the Galactic bulge, including $\sim 300$
lensing events (Wo\'zniak 2000). Of these, two or three appear to show 
demagnification preceding or following the central peak in brightness,
although the statistical significance of the demagnification has not 
yet been established (Wo\'zniak, private communication). Demagnification 
cannot be produced by gravitational lensing by one or more point masses,
 but would be a characteristic signature of gaseous lensing.

\section{Summary}

In this paper we have considered in detail the lensing of stars in the LMC
by small self-gravitating molecular clouds which 
have been proposed as a candidate for the  dark matter
in the Galaxy.
Lensing would occur because of the refraction of optical light in the clouds,
with resulting  magnification of the source.
We have developed a semianalytical formalism for calculation of the 
lensing rate including the spatial distribution of the dark
matter, finite velocity dispersions of the lensing clouds and
the source stars in LMC, and proper motion of the observer and LMC
itself. 

Our calculations were carried out for a single mass and size cloud population.
One can easily extend the analysis to a
distribution of cloud masses and sizes by simply convolving event rates 
obtained for a uniform cloud population with
the appropriate distributions.
 
Lensing events might be detectable by  
searches for gravitational microlensing. This has 
allowed us to obtain constraints on the cloud properties
by comparing our calculations with observational results  obtained 
by the MACHO
collaboration.

We found that almost
the only possibility for the dark matter to be in the form of 
such molecular clouds is for the clouds 
to be sufficiently weak lenses so that their lensing effects
are below the detection threshold of the MACHO experiment,
since otherwise a very large gaseous lensing event rate
would have been detected. This still
leaves an allowed region in  parameter space 
 where these clouds could exist and not contradict the  limitations 
posed by lensing experiments and  simple physical considerations.
 
Though we have performed event rate calculations for only one 
halo model [given by equation (\ref{halo})], 
we expect
our results to be relatively insensitive to the  particular form 
of the spatial distribution of these clouds, or the assumed 
shape of the velocity distribution of the clouds and the stars in the LMC. 

Future microlensing experiments with a lower detection  threshold 
magnification $A_t$
could either detect these clouds or strengthen the constraints
on their properties.

\section{Acknowledgements} 

We thank B.Paczy\'nski and P.Wo\'zniak for very helpful discussions, and
  R.H.Lupton for availability of the SM plotting package.
This research has been supported in part by NSF grants
AST-9619429 and AST-9988126.

\appendix

\section{Derivation of equations (\ref{rate-tot}) -
(\ref{rate-t_min})}\label{A1}

For the purposes of the lensing rate calculations 
the velocities of the lens, observer, and source along the line of sight
are not important, because only transverse motions are significant.
It means that we can immediately perform integrations in 
equation (\ref{rate-raw})  over $v_{S z}$ and $v_{Lz}$.

Taking into account (\ref{timescale}) we can perform integration 
over $\varphi$
 in (\ref{rate-raw})
and define the result  as a function $\kappa({\bf v}_{\perp})$: 
\begin{equation}
\kappa({\bf v}_{\perp})\equiv\frac{1}{\delta(x,A_t)} 
\int\limits^{\delta(x,A_t)}_0
\phi_{\tau}(\tau(\varphi))d\varphi=
\int\limits^1_0
\phi_{\tau}
\left(\frac{2\delta(x,A_t) D_{OL}}{|{\bf v}_{\perp}|}\sqrt{1-s^2}\right)ds,
\label{kappa}
\end{equation}
so that now
\begin{equation}
d\dot N=2\delta\frac{\Sigma_S}{D_{OL}}
\int\limits^{\infty}_{-\infty} dv_{L x}
\int\limits^{\infty}_{-\infty} dv_{L y}
\int\limits^{\infty}_{-\infty} dv_{S x}
\int\limits^{\infty}_{-\infty} dv_{S y}
|{\bf v}_{\perp}| f_L(v_{L x},v_{L y})f_S(v_{S x},v_{S y})
\kappa(|{\bf v}_{\perp}|).
\label{rate-1}
\end{equation}
We now change  variables from 
$v_{L x}$ and $v_{L y}$ to $v_x$ and $v_y$ via  equation
(\ref{vels}):
\begin{eqnarray}
d\dot N=2\delta\frac{d\Sigma_L}{\pi^2 \sigma_L^2 \sigma_S^2 D_{OL}}
\int\limits^{\infty}_{-\infty} dv_x
\int\limits^{\infty}_{-\infty} dv_y
\int\limits^{\infty}_{-\infty} dv_{S x}
\int\limits^{\infty}_{-\infty} dv_{S y}
\sqrt{v_x^2+v_y^2}\kappa(\sqrt{v_x^2+v_y^2})\nonumber\\
\times\exp\left\{
-\frac{(v_x-(1-x)v_{O x}-x v_{S x})^2+(v_y-x v_{S y})^2}
{\sigma_L^2}-\frac{(v_{S x}-v_{c x})^2+(v_{S y}-v_{c y})^2}
{\sigma_S^2}
\right\}.
\label{rate-2}
\end{eqnarray}

We integrate over $v_{S x}$ and $v_{S y}$
to obtain after lengthy but straightforward calculations
\begin{eqnarray}
d\dot N=2\delta\frac{\Sigma_S}
{\pi (\sigma_L^2+x^2\sigma_S^2) D_{OL}}
\int\limits^{\infty}_{-\infty} dv_x
\int\limits^{\infty}_{-\infty} dv_y
\sqrt{v_x^2+v_y^2}\kappa(\sqrt{v_x^2+v_y^2})\nonumber\\
\times\exp\left\{
-\frac{v_x^2+v_y^2}{\sigma_L^2+x^2\sigma_S^2}+\zeta_1 v_x+
\zeta_2 v_y-C
\right\},
\label{rate-2.5}
\end{eqnarray}
where
\begin{equation}
C=\frac{x^2 v_{c y}^2 + (x v_{c x}+(1-x)v_{O x})^2}{\sigma_L^2+x^2\sigma_S^2},
\label{C}
\end{equation}
\begin{equation}
\zeta_1=2\frac{x v_{c x}+(1-x)v_{O x}}{\sigma_L^2+x^2\sigma_S^2},~~~~~~~~~~
\zeta_2=2\frac{x v_{c y}}{\sigma_L^2+x^2\sigma_S^2}.
\end{equation}

Introducing  polar coordinates  in velocity space 
$v_x=v\cos\beta$, $v_y=v\sin\beta$ and normalizing $v$ to some characteristic
velocity $v_{ch}$ we get
\begin{eqnarray}
d\dot N=2\delta\frac{\Sigma_S v_{ch}^3}
{\pi (\sigma_L^2+x^2\sigma_S^2) D_{OL}}e^{-C}
\int\limits^{\infty}_0 du
~u^2\kappa(uv_{ch})e^{-A u^2}
\int\limits^{2\pi}_0 e^{u v_{ch}(\zeta_1\cos\beta+\zeta_2\sin\beta)}d\beta,
\label{rate-3}
\end{eqnarray}
where
\begin{equation}
A=\frac{v_{ch}^2}{\sigma_L^2+x^2\sigma_S^2}.
\label{A}
\end{equation}
The last integral in equation (\ref{rate-3}) can be reduced to
\begin{eqnarray}
\int\limits^{2\pi}_0 e^{u v_{ch}
\sqrt{\zeta_1^2+\zeta_2^2}cos(\beta+\beta_0)}d\beta=
\int\limits^{2\pi}_0 e^{u v_{ch}\sqrt{\zeta_1^2+\zeta_2^2}cos\beta}d\beta=
2\pi I_0\left(B u\right),
\label{bess}
\end{eqnarray}
where $I_0$ is the modified Bessel function of order zero, and
\begin{equation}
B=v_{ch}\sqrt{\zeta_1^2+\zeta_2^2}=\frac{2v_{ch}}{\sigma_L^2+x^2\sigma_S^2}
\sqrt{x^2 v_{c y}^2 + (x v_{c x}+(1-x)v_{O x})^2}=2\sqrt{AC}.
\label{B}
\end{equation} 

So, finally we get for the event rate produced by one lens
\begin{eqnarray}
d\dot N(x)=4\delta\frac{\Sigma_S v_{ch}^3}
{(\sigma_L^2+x^2\sigma_S^2) D_{OL}}e^{-C}
\int\limits^{\infty}_0 
u^2\kappa(uv_{ch})e^{-A u^2}I_0\left(B u\right)du.
\label{rate}
\end{eqnarray}

Thus, the  total
event rate per source, produced by all lenses between us and LMC, is
given by formula (\ref{rate-tot}).

Now, we obtain the formula for the timescale distribution 
(\ref{dist}).
Note that $\tau$ and $\varphi$ -- the angular distance 
from the lens's trajectory 
to the source star in the perpendicular direction -- are directly 
related by equation (\ref{timescale}):
\begin{equation}
d\varphi=-\delta(x,A_t)\frac{d\tau}{\tau}\frac{u^2}{\sqrt{1-u^2}},
\end{equation}
with $u=v_{\perp}/v_{ch}$, $v_{ch}=2b_{0 t}(x,A_t)/\tau$. To obtain
$d\dot N_{tot}/d\tau$ we simply omit the integration over
$\varphi$ in (\ref{rate-raw}). Also, the velocity $v_{\perp}$
is limited by $2b_{0 t}(x,A_t)/\tau$.
All the other steps are analogous to
those which we have done in derivation of (\ref{rate-tot}) and we obtain
 equation (\ref{dist}).

If we are interested only in events whose durations are larger than 
some chosen value $\tau_{min}$, then to get the total rate we should
integrate in (\ref{rate-raw}) over $\varphi$ not up to $\delta(x,A_t)$, but
up to 
\begin{equation}
\varphi_{max}=\sqrt{\delta^2-\frac{v_{\perp}^2\tau_{min}^2}{4 D_{OL}}}=
\delta\sqrt{1-u^2},
\end{equation}
with $u=v_{\perp}\tau/2\delta x D$. 
One can easily see that 
we will get in this case for $\dot N_{tot}(\tau>\tau_{min})$  the formula
(\ref{rate-t_min}).

\section{Detection efficiency function}\label{A2}

The detection efficiency function 
$\phi$ characterizes the sensitivity of the 
experiment to  events with different durations. We will use
the detection efficiency $\phi_{\hat t}(A,\hat t)$ 
assumed by MACHO experiment 
to estimate how it would affect gaseous lensing (see Figure 8 in
Alcock et al. (1997)).

We first recall some simple 
facts about gravitational microlensing.
The amplification $M$ in the case of gravitational microlensing is
determined by the formula
\begin{eqnarray}
M(u)=\frac{u^2+2}{u(u^2+4)^{1/2}}, ~~~~
u=(u_{min}^2+4t^2/\hat t^2)^{1/2}\nonumber\\
u_{min}=\frac{b_{min}}{r_E},~~~~\hat t=2\frac{r_E}{v_{\perp}},~~~~
r_E=\left[\frac{4GmDx(1-x)}{c^2}\right]^{1/2},
\label{gml}
\end{eqnarray}
where $b_{min}$ is the impact parameter, $m$ is the mass of the lensing 
object, $t$ is the time of observation,
and $v_{\perp}$ is, again, the relative velocity of the lens and the 
source star. 

Recall that we defined the timescale $\tau$ to be the time spent with
magnification $M>A_t$. From equation (\ref{gml}) we find that, for
gravitational microlensing,
\begin{eqnarray}
\tau_{gm}=\sqrt{2}\hat t\left[\frac{A_t}{\sqrt{A_t^2-1}}-
\frac{A_{max}}{\sqrt{A_{max}^2-1}}
\right]^{1/2}\nonumber   \\=
\hat t\left[u_t^2-u_{min}^2
\right]^{1/2},
\label{tau-b}
\end{eqnarray}
where $u_t$ is the value of $u$ at which the magnification is equal 
to $A_t$ and $A_{max}$ is the maximum magnification for a given 
$u_{min}$. 

We use a very simple algorithm for converting detection 
efficiency function from $\hat t$ to
$\tau$: since $\tau \propto \hat t$, their ratio $\tau/\hat t$ is
a function of $A_{max}$ (or $u_{min}$) for a given $A_t$.
Thus we can average this ratio over the distribution of the 
maximum magnifications $A_{max}$ and assume that the value
of known $\phi_{\hat t}(\hat t)$ is equal to the 
$\phi_{\tau}(<\tau>)$, that is
\begin{equation}
\phi_{\tau}(\tau)=\phi_{\hat t}\left(\tau\left(
<\tau/\hat t>\right)^{-1}\right).
\end{equation}

The calculation of $<\tau/\hat t>$ is quite straightforward and
can be most easily done in terms of $u_{min}$ (see the last equality in
equation (\ref{tau-b})). Since the distribution  of lensing events
over $u_{min}$ is just constant 
\begin{equation}
<\tau/\hat t>=\frac{\int\limits^{u_t}_0\left[u_t^2-u_{min}^2
\right]^{1/2}du_{min}}{\int\limits^{u_t}_0 du_{min}}=
\frac{\pi}{4}u_t=\frac{\pi}{2\sqrt{2}}
\left(\frac{A_t}{\sqrt{A_t^2-1}}-1\right)^{1/2}.
\end{equation}
For $A_t=1.35$, accepted by MACHO experiment, $<\tau/\hat t>=
0.776$, which means that 
$\phi_{\tau}(\tau)=\phi_{\hat t}(1.288\tau)$.

\end{document}